\documentclass[runningheads]{llncs}
\usepackage[T1]{fontenc}
\usepackage{graphicx}
\usepackage{xcolor}
\usepackage[colorinlistoftodos,textsize=small]{todonotes}
\usepackage{amsmath}
\usepackage{booktabs} 
\usepackage{amssymb}
\usepackage{svg}
\usepackage{subcaption}
\usepackage{float}
\usepackage{hyperref}
\usepackage{makecell}

\begin{document}
%
\title{A Biophysically-Conditioned Generative Framework for 3D Brain Tumor MRI Synthesis}
\titlerunning{Biophysically Generative Framework for Brain Tumor MRI Synthesis}
%
\author{Valentin Biller \inst{1}\and
Lucas Zimmer \inst{1,2} \and
Ayhan Can Erdur \inst{1} \and
Sandeep Nagar \inst{1} \and
Daniel Rückert \inst{1,2,3} \and
Niklas Bubeck \inst{1,2}* \and
Jonas Weidner \inst{1,2}* 
}
%
\authorrunning{V. Biller et al.}
%
\institute{
Technical University of Munich, Germany. \and
Munich Center for Machine Learning (MCML)\and
Imperial College London\\
* equal contribution \\
\email{\{valentin.biller, niklas.bubeck, j.weidner\}@tum.de} 
}
\maketitle              
\begin{abstract}
Magnetic resonance imaging (MRI) inpainting supports numerous clinical and research applications. We introduce the first generative model that conditions on voxel‑level, continuous tumor concentrations to synthesize high-fidelity brain tumor MRIs. For the BraTS 2025 Inpainting Challenge\footnote{Profil-ID:3504517}, we adapt this architecture to the complementary task of healthy tissue restoration by setting the tumor concentrations to zero. Our latent diffusion model conditioned on both tissue segmentations and the tumor concentrations generates 3D spatially coherent and anatomically consistent images for both tumor synthesis and healthy tissue inpainting. For healthy inpainting, we achieve a PSNR of $18.5$, and for tumor inpainting, we achieve $17.4$.
Our code is available at: \url{https://github.com/valentin-biller/ldm.git}
\end{abstract}
\section{Introduction}
A wide range of automated analysis tools for brain MRI is available for clinical decision support. However, these tools often assume healthy anatomy, limiting their reliability when applied to pathological images. For brain tumor patients, this mismatch is particularly relevant, as MRI typically commences post-diagnosis when lesions are already present. This limits the effectiveness of algorithms that depend on healthy anatomical priors, such as parcellation, tissue segmentation, or brain extraction.
Data-driven generative models can synthesize healthy tissue by inpainting or reconstructing resection cavities while preserving surrounding anatomical structures. These models can also be leveraged to generate anatomically consistent tumorous images. To generate realistic healthy tissue, establishing a method for tumor synthesis supports a unified framework that benefits from learning both tasks. As tumor synthesis has been studied more extensively, it serves as a logical starting point for reviewing prior work before addressing inpainting as a downstream application.

\noindent Early efforts in tumor image synthesis relied on 2D generation or the use of convolutional architectures \cite{wolleb2022swiss,truong2024synthesizing}. Diffusion models have recently emerged as state-of-the-art for high-fidelity image synthesis. Operating the diffusion process in a compressed latent space further improves efficiency and memory footprint \cite{rombach2022latent}. While most medical implementations remain restricted to 2D slices, recent work demonstrates fully 3D diffusion for volumetric CT and MRI \cite{bohnenberger2024diffusion}. Building on these advances, we employ a 3D latent diffusion framework that preserves inter-slice consistency, avoiding the misalignment artifacts reported for slice-wise approaches \cite{liu2025treatment}.

\noindent Most generative approaches for brain tumors rely on discrete tumor segmentation masks \cite{dorjsembe2024conditional,truong2024synthesizing}. While effective for defining gross tumor geometry, these masks lack biological realism, treating tumor regions as uniformly dense and failing to capture infiltrative growth. Since gliomas often extend beyond MRI-visible margins, biophysical tumor growth modeling provides a means to reveal this hidden infiltration and reduces reliance on standard uniform treatment margins, which do not account for patient-specific tumor spread \cite{weidner2024spatial,balcerak2025individualizing}. By conditioning our models on continuous tumor concentrations generated through such growth modeling, we overcome the limitations of discrete masks \cite{bortfeld2022modeling,balcerak2025individualizing,balcerak2024physics,weidner2024spatial,weidner2024learnable,ezhov2023learn}. These scalar fields encode spatially varying tumor density, capturing both the visible tumor bulk and subtle infiltration into surrounding tissue, while providing fine-grained control over lesion appearance. By leveraging biophysically grounded priors, our approach better aligns with clinical reality.

\noindent We evaluate our method on the Brain MR Image Inpainting Challenge \cite{kofler2023brain} by setting the tumor concentration to zero, effectively asking the model to remove the lesion. Quantitative and qualitative results confirm that the same network can act as a controllable tumor generator and an anatomically aware inpainting tool.

\noindent Our contributions are threefold:
\begin{itemize}
    \item We developed the first 3D generative brain tumor MRI model, which is conditioned on continuous tumor concentrations generated by biophysical tumor growth models.
    \item We show that the same model excels at healthy brain inpainting under the zeroed condition.
    \item Our pipeline can easily be extended to further modalities like different MRI sequences or PET.
\end{itemize}
\section{Method}
\begin{figure}
\begin{center}
\includegraphics[width=0.8\textwidth]{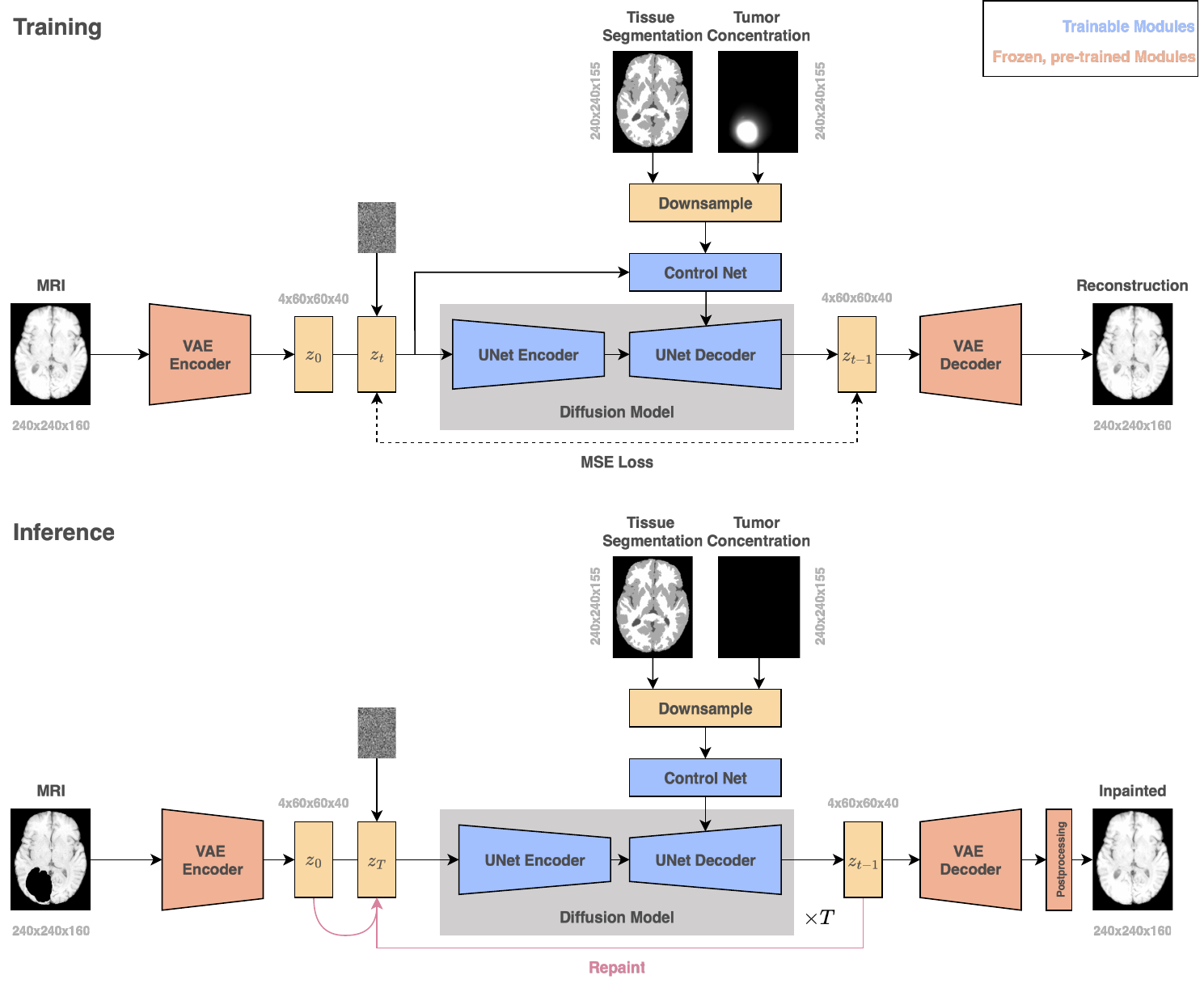}
\caption{For the training of our model, we input the tissue segmentations and tumor concentrations as conditions to the latent diffusion model. During inference, we set the tumor concentration to 0 to inpaint the voided regions as healthy brain tissue.} \label{fig:overview} 
\end{center}
\end{figure}

Our proposed method, illustrated in Figure~\ref{fig:overview}, is a two-stage 3D latent diffusion framework designed for anatomically consistent brain tumor MRI synthesis and inpainting. We condition a latent diffusion model on both the tissue segmentations and a continuous tumor concentration, which encodes spatially varying tumor cell density and is generated by a biophysical growth model. This scalar field enables fine-grained control over lesion appearance, allowing the model to synthesize realistic tumor-bearing images or, by setting the concentration to zero, perform healthy tissue inpainting.

\subsection{Overview}
\label{sec:overview}
\noindent\textit{\textbf{Autoencoder.}} 
To efficiently represent high-resolution 3D brain MR images, we adopt the pretrained variational autoencoder (VAE) from the MAISI framework~\cite{guo2025maisi} to map input T1-weighted volumes into a compact latent space. Formally, given an input image volume $\mathbf{x} \in \mathbb{R}^{240 \times 240 \times 155}$, the encoder network $E_{\phi}$ produces a latent distribution $q_{\phi}(\mathbf{z}|\mathbf{x})$, parameterized by a mean and variance, from which a latent code $\mathbf{z} \in \mathbb{R}^{4 \times 60 \times 60 \times 40}$ is sampled via the reparameterization trick. This compression reduces the spatial resolution by a factor of $4$ in each dimension while preserving semantically relevant anatomical features. The decoder network $D_{\theta}$ reconstructs the image from the latent representation, yielding $\hat{\mathbf{x}} = D_{\theta}(\mathbf{z})$. The VAE used in our pipeline is a pretrained model, trained independently of the inpainting task. Its parameters $(\phi, \theta)$ are kept fixed throughout all stages of our method. This latent encoding significantly reduces computational cost and memory footprint during training and inference.

\noindent\textit{\textbf{Latent Diffusion.}} 
We employ a generative model based on the Denoising Diffusion Probabilistic Models (DDPM) framework~\cite{ho2020denoising}, which learns to reverse a forward diffusion process defined over latent variables. The forward process progressively perturbs a clean latent sample $z_0$ into a sequence of noisy versions $\{z_t\}_{t=1}^T$ according to the marginal distribution:
\begin{equation}
    q(z_t \mid z_0) = \mathcal{N}(\sqrt{\bar\alpha_t}z_0, (1 - \bar\alpha_t)\mathbf{I}),
\end{equation}
where $\{\bar\alpha_t\}_{t=1}^T$ is the level of preserved signal. A 3D U-Net $\epsilon_\theta(z_t, t)$ is trained to estimate the noise component $\epsilon$ added at each timestep, minimizing the following objective:
\begin{equation}
    \mathcal{L}_{\text{gen}}(z) = \mathbb{E}_{z_0, \epsilon, t} \left[ \| \epsilon_\theta(z_t, t) - \epsilon \|^2_2 \right].
\end{equation}

\noindent\textit{\textbf{Tissue and Tumor Conditioning.}} 
Anatomical conditioning is provided via one-hot encoded tissue segmentations for cerebrospinal fluid (CSF), gray matter (GM), and white matter (WM), obtained through atlas-based registration using the \texttt{gbm\_bench}\footnote{\url{github.com/LMZimmer/gbm_bench}} framework. Additionally, a continuous scalar field denoting voxel-wise tumor concentrations in the range $[0, 1]$ is used to encode relative tumor cell density, derived from a biophysical growth model~\cite{weidner2024spatial}. For inpainting, these tumor concentrations are set to zero to indicate lesion absence. All conditioning inputs are downsampled to the latent resolution via nearest-neighbor interpolation. Conditioning is implemented following a ControlNet-style architecture~\cite{zhang2023adding}, where the concatenated tissue segmentations and tumor concentrations are processed by a separate convolutional branch and fused with the diffusion U-Net through feature-wise addition at multiple layers, enabling structured spatial guidance during generation. This design enables the model to synthesize tumors in accordance with the provided tumor concentrations while ensuring that the surrounding anatomy remains consistent with the underlying tissue segmentations.

\subsection{Inference}
To leverage the strong theoretical foundations and principled probabilistic modeling, we adopt the Denoising Diffusion Probabilistic Models (DDPM) sampling strategy~\cite{ho2020denoising} in the latent space. 
The latent update at timestep $t-1$ is given by:
\begin{align}\label{equation:reverse_diffusion_temp}
    z_{t-1} =\ & \sqrt{\bar\alpha_{t-1}} \left( \frac{x_t - \sqrt{1 - \bar\alpha_t} \epsilon_\theta(z_t, t, c^*)}{\sqrt{\bar\alpha_t}} \right) \nonumber \\
    & + \sqrt{1 - \bar\alpha_{t-1} - \sigma_t^2} \cdot \epsilon_\theta(z_t, t, c^*) + \sigma_t \epsilon_t, 
\end{align}
where $c^*$ corresponds to the conditioning information encoding anatomical tissue segmentations and tumor concentrations, \( \epsilon_t \sim \mathcal{N}(0, \mathbf{I}) \) is noise and \( \sigma_t \) is the timestep-dependent variance. 

\noindent\textit{\textbf{Known Region Injection for Inpainting.}} 
In the inpainting task, reconstruction is guided through a process commonly referred to as known region injection. At each diffusion timestep $t$, voxels corresponding to regions with known ground truth values are injected back into the model's current denoised latent estimate to enforce spatial consistency. Formally, the denoised latent $ \hat{z}_t $ at timestep $t$ is partially overwritten by a noisy version of the ground truth latent, where the noise level matches the current timestep. This noisy ground truth latent $z_t^{GT}$ is sampled from the forward diffusion process as
\[
z_t^{GT} \sim \mathcal{N}\left(\sqrt{\bar\alpha_t} z_0^{GT}, (1 - \bar\alpha_t) \mathbf{I}\right),
\]
with $z_0^{GT}$ representing the clean ground truth latent. The injection is implemented as
\[
\hat{z}_t \leftarrow M \odot z_t^{GT} + (1 - M) \odot \hat{z}_t,
\]
where $M$ is a binary mask indicating known voxel locations and $\odot$ denotes element-wise multiplication. This procedure ensures that the known regions remain consistent with the original data distribution throughout the reverse diffusion process, while allowing the model to synthesize plausible content in the unknown regions. As the noise level decreases over timesteps, the injected regions converge to their true values, thereby facilitating accurate and spatially coherent inpainting of the missing tissue. Notably, the mask used during this process was a slightly expanded version of the unknown region, obtained by applying one iteration of binary dilation to $1 - M$. This helps ensure a smoother transition between known and unknown regions, supporting more realistic tissue reconstruction.

\noindent\textit{\textbf{Repainting Mechanism.}}  
While known region injection enforces consistency, it introduces discontinuities at the boundary between known and unknown regions. This naïve injection causes disharmony, as the model cannot smoothly blend generated and fixed content - it lacks visibility into how its outputs interact with the static known regions over time. To address this, we use the RePaint algorithm by Lugmayr et al.~\cite{lugmayr2022repaint} that refines boundary regions through targeted re-noising and resampling. 
In each denoising step, the content for the known region ($x_{t-1}^{\text{known}}$) is sampled using the known pixels in the given image $m \odot x_0$, while the content for the unknown region ($x_{t-1}^{\text{unknown}}$) is sampled from the model, given the previous iteration $x_t$. These components are then composited using a binary mask $m$ to form the complete latent for the next step, as described by the equation:
\[
x_{t-1} = m \odot x_{t-1}^{\text{known}} + (1-m) \odot x_{t-1}^{\text{unknown}}
\]
To improve harmony between these regions, a resampling technique is used, which involves taking steps both backward and forward in diffusion time. This allows the model to iteratively re-contextualize and harmonize the generated content with the known image information, improving overall coherence and semantic plausibility without disturbing known regions.

\noindent\textit{\textbf{Image-Space Postprocessing.}}
To improve visual coherence at the boundaries between inpainted and known regions, we apply image-space postprocessing composed of poisson blending~\cite{perez2023poisson} and histogram equalization~\cite{kim1997contrast}. Poisson blending refines low-level transitions by harmonizing gradient fields across region boundaries, mitigating visible seams caused by pixel discontinuities. Complementarily, histogram equalization aligns intensity distributions between the synthesized region and the known context. This is computed using non-black voxels from both the generated output and the corresponding ground truth, ensuring normalization focuses on anatomically relevant structures. Together, these techniques enhance both the perceptual smoothness and photometric consistency of the inpainted images.
\begin{figure}[t]
\includegraphics[width=\textwidth]{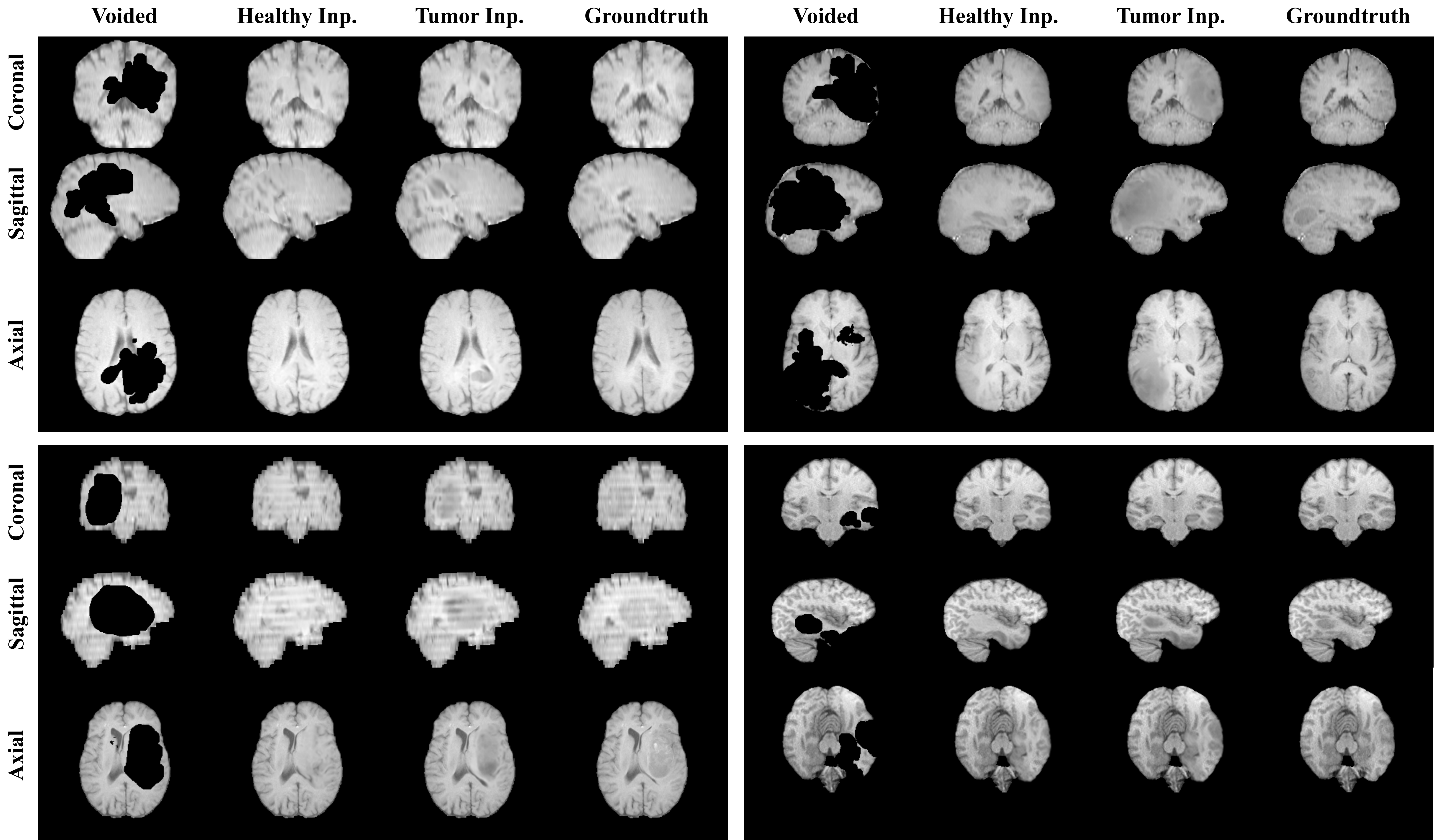}
\caption{Qualitative inpainting results are presented for four representative subjects across the coronal, sagittal, and axial views. We showcase our ability to reconstruct missing regions in a 3D spatially coherent and anatomically consistent manner, both for tumor reconstruction and healthy tissue, with a single model.} \label{fig:example_healthy_inpainting}
\end{figure}

\section{Experiments}
\noindent\textit{\textbf{Dataset.}} 
We employed the publicly available BraTS 2021 dataset~\cite{baid2021brats} in conjunction with several additional private and public datasets (ucsf-pdgm, tcga-gbm, tcga-lgg, Rembrandt)\cite{calabrese2022ucsf_pdgm,scarpace2016tcga_gbm,pedano2016tcga_lgg,scarpace2019rembrandt}, collectively comprising MRI scans of brain tumor patients. The aggregated cohort consists of 3,602 subjects, partitioned into 80\% for training (2,881 subjects) and 20\% for validation (721 subjects). All volumes were spatially normalized via co-registration to a standardized anatomical template, resampled to an isotropic voxel resolution of 1\,mm$^3$, and subjected to skull-stripping to remove non-brain tissues. For the purpose of this study, only the T1-weighted MRI modality was utilized. Each volume was intensity-normalized to the range $[0,1]$, zero-padded to a uniform size of $240 \times 240 \times 160$, and subsequently cropped back to $240 \times 240 \times 155$ post-generation to preserve the original brain region.

\noindent\textit{\textbf{Implementation Details.}} 
The diffusion model is trained with a linear noise schedule, where $\beta_t$ increases from $1 \times 10^{-4}$ to $0.02$ over 1,000 timesteps. For inference, we adopt the sampling schedule from the RePaint algorithm by Lugmayr et al.~\cite{lugmayr2022repaint}, which uses 250 timesteps and incorporates their resampling strategy with a jump length of 10 and 10 resampling steps. The model architecture is implemented using the MONAI framework~\cite{cardoso2022monai}.

\noindent\textit{\textbf{Training Details.}} 
Training was conducted using PyTorch Lightning on 2 NVIDIA H100 GPUs (94 GB each) with a batch size of 2 over a duration of approximately 2.5 weeks. Optimization employed the AdamW algorithm with weight decay of 0.01 and an initial learning rate of $1 \times 10^{-4}$, modulated by a cosine annealing scheduler. The noise prediction U-Net and the ControlNet modules were trained jointly differing from the original ControlNet scheme. 



\noindent\textit{\textbf{Healthy Tissue Inpainting.}}
To evaluate the performance of tissue inpainting, we adopted the BraTS-based inpainting dataset generation protocol\footnote{\url{github.com/BraTS-inpainting/2023_challenge}}. For each subject, the volumetric MRI data were masked in two regions: one containing the tumor and another selected randomly from healthy tissue, simulating missing regions. To rigorously assess inpainting accuracy, quantitative performance metrics - including SSIM, PSNR, MAE, MSE, RMSE, and MSLE - were computed exclusively within the masked healthy region, as ground truth data are available only for that area.

\noindent\textbf{\textit{Tumor Inpainting.}}
The same dataset and masking protocol were used for evaluating tumor inpainting. In this case, performance metrics were computed in both the healthy and tumorous regions, as ground truth information is available for both areas. Unlike in the healthy tissue evaluation, the tumor concentrations were not zeroed out but retained as provided in the dataset. A detailed description of this representation is provided in Section~\ref{sec:overview}.
\begin{figure}[p] 
\centering

\begin{minipage}{\textwidth}
\centering
\captionof{table}{Quantitative values for both healthy tissue (a) and tumor (b) inpainting.}
\label{tab:metrics}
\vspace{0.5em}
\begin{minipage}{0.48\textwidth}
\centering
\begin{tabular}{lccc}
\toprule
\textbf{Metric} & \textbf{Mean} & \textbf{Median} & \textbf{Std} \\
\midrule
SSIM $\uparrow$ & 0.754 & 0.746 & 0.134 \\
PSNR $\uparrow$ & 18.542 & 18.140 & 3.121 \\
MAE $\downarrow$ & 0.088 & 0.084 & 0.032 \\
MSE $\downarrow$ & 0.017 & 0.015 & 0.011 \\
RMSE $\downarrow$ & 0.123 & 0.121 & 0.040 \\
MSLE $\downarrow$ & 0.007 & 0.006 & 0.005 \\
\bottomrule
\end{tabular}
\caption*{(a) Healthy Tissue Inpainting}
\end{minipage}
\hfill
\begin{minipage}{0.48\textwidth}
\centering
\begin{tabular}{lccc}
\toprule
\textbf{Metric} & \textbf{Mean} & \textbf{Median} & \textbf{Std} \\
\midrule
SSIM $\uparrow$  & 0.578 & 0.576 & 0.090 \\
PSNR $\uparrow$  & 17.360 & 17.664 & 2.262 \\
MAE $\downarrow$ & 0.104 & 0.095 & 0.041 \\
MSE $\downarrow$ & 0.022 & 0.017 & 0.024 \\
RMSE $\downarrow$ & 0.141 & 0.131 & 0.047 \\
MSLE $\downarrow$ & 0.009 & 0.007 & 0.011 \\
\bottomrule
\end{tabular}
\caption*{(b) Tumor Inpainting}
\end{minipage}
\end{minipage}

\vspace{1.5em}

\begin{minipage}{\textwidth}
\centering
\begin{minipage}{0.48\textwidth}
\centering
\includegraphics[width=\textwidth]{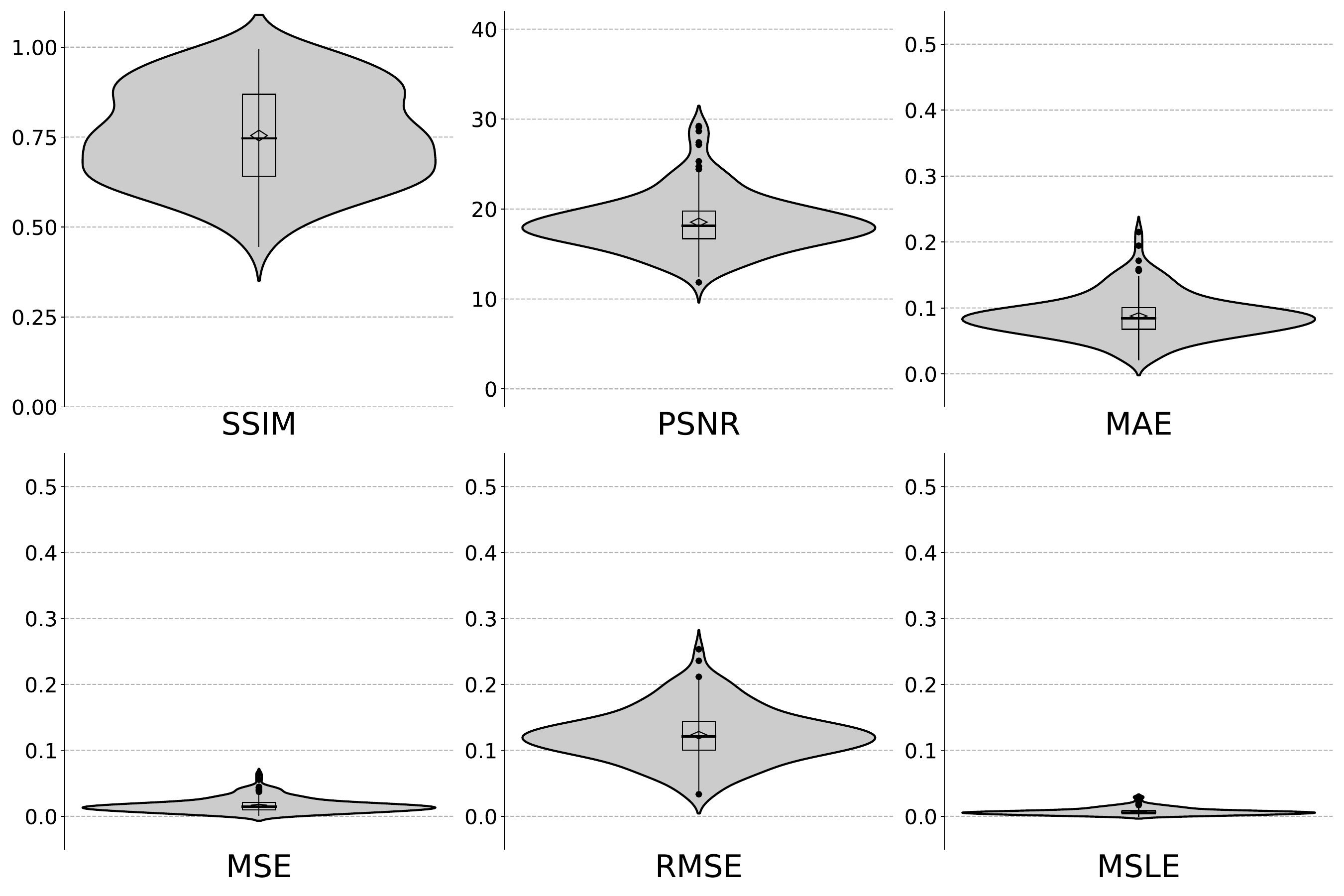}
\caption*{(a) Healthy Tissue Inpainting}
\end{minipage}
\hfill
\begin{minipage}{0.48\textwidth}
\centering
\includegraphics[width=\textwidth]{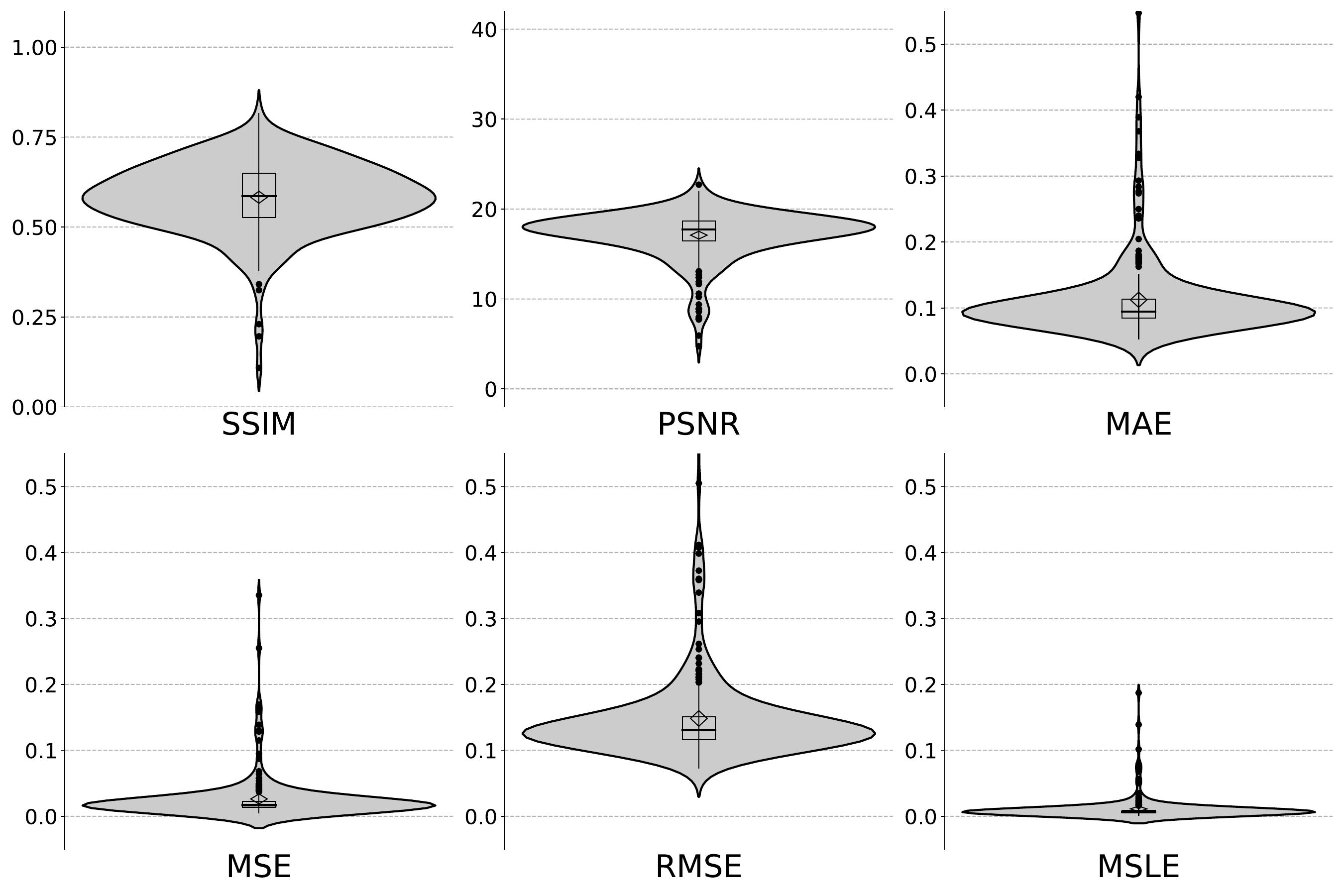}
\caption*{(b) Tumor Inpainting}
\end{minipage}
\caption{Violin plots of quantitative metrics for healthy (a) and tumor (b) inpainting. Median performance is indicated by a thick horizontal line, mean performance by a rhombus. The box bounds represent the first and third quartiles, and density is shown by the violin plot.}
\label{fig:metrics_combined}
\end{minipage}

\vspace{3em}

\captionof{table}{Ablation study results for both healthy tissue (a) and tumor (b) inpainting comparing no postprocessing (I), histogram equalization (HE), and poisson blending (PB).}
\label{tab:ablation}

\vspace{1.5em}
    
\begin{minipage}{\textwidth}
\centering
\begin{tabular}{ccc @{\quad} cccccc}
\toprule
\textbf{I} & \textbf{HE} & \textbf{PB} & \textbf{SSIM $\uparrow$} & \textbf{PSNR $\uparrow$} & \textbf{MAE $\downarrow$} & \textbf{MSE $\downarrow$} & \textbf{RMSE $\downarrow$} & \textbf{MSLE $\downarrow$} \\
\midrule
\checkmark & & & 0.715 & 14.615 & 0.153 & 0.045 & 0.198 & 0.016 \\
\checkmark & \checkmark & & 0.735 & 17.514 & 0.097 & 0.021 & 0.138 & 0.009 \\
\checkmark & \checkmark & \checkmark & 0.754 & 18.542 & 0.088 & 0.017 & 0.123 & 0.007 \\
\bottomrule
\end{tabular}
\caption*{(a) Healthy Tissue Inpainting}
\end{minipage}

\vspace{1.5em}

\begin{minipage}{\textwidth}
\centering
\begin{tabular}{ccc @{\quad} cccccc}
\toprule
\textbf{I} & \textbf{HE} & \textbf{PB} & \textbf{SSIM $\uparrow$} & \textbf{PSNR $\uparrow$} & \textbf{MAE $\downarrow$} & \textbf{MSE $\downarrow$} & \textbf{RMSE $\downarrow$} & \textbf{MSLE $\downarrow$} \\
\midrule
\checkmark & & & 0.549 & 13.864 & 0.175 & 0.054 & 0.217 & 0.019 \\
\checkmark & \checkmark & & 0.555 & 16.767 & 0.110 & 0.025 & 0.151 & 0.010 \\
\checkmark & \checkmark & \checkmark & 0.578 & 17.360 & 0.104 & 0.022 & 0.141 & 0.009 \\
\bottomrule
\end{tabular}
\caption*{(b) Tumor Inpainting}
\end{minipage}

\end{figure}

\section{Results}

\noindent\textbf{\textit{Quantitative Results.}}
The quantitative metrics reported in Table~\ref{tab:metrics} demonstrate that the proposed inpainting model achieves moderate to high performance across structural, perceptual, and reconstruction-based evaluation criteria for both healthy tissue and tumor inpainting tasks. The distribution of these metrics, visualized through violin plots in Figure~\ref{fig:metrics_combined}, reveals a concentrated central tendency with a pronounced tail of lower-performing cases—particularly in SSIM and RMSE. These outliers suggest the presence of challenging anatomical or masking conditions to which the model may be particularly sensitive, highlighting the need for further stratified or case-specific analysis. In Table \ref{tab:challenge}, we show our results at the BraTS challenge. 

\noindent\textbf{\textit{Qualitative Results.}}
The qualitative examples shown in Figure~\ref{fig:example_healthy_inpainting} illustrate that the proposed inpainting model is capable of generating anatomically plausible reconstructions, exhibiting spatial coherence across coronal, sagittal, and axial planes. The reconstructed regions generally preserve structural continuity and align well with surrounding anatomical features. However, in certain cases, subtle texture inconsistencies and imperfect transitions between the original and inpainted regions are observable, particularly near region boundaries. These artifacts suggest limitations in the model’s ability to fully harmonize edge details, indicating potential areas for improvement in boundary refinement and texture blending mechanisms.

\noindent\textbf{\textit{Ablation Study.}}
An ablation study, detailed in Table~\ref{tab:ablation}, was conducted to evaluate the contribution of individual postprocessing components on the final output. The results clearly demonstrate the efficacy of a sequential enhancement pipeline. The baseline model without any postprocessing yields the lowest performance across all metrics. The introduction of histogram equalization provides a substantial improvement, most notably increasing the PSNR from 14.615 to 17.514 and reducing the MAE from 0.153 to 0.097. The subsequent application of poisson blending provides a further, albeit more modest, refinement, improving the PSNR to 18.542 and the MAE to 0.088. This incremental enhancement underscores the value of both steps: histogram equalization is critical for correcting the overall intensity distribution, while poisson blending is effective in seamlessly integrating the inpainted patch, which directly addresses the boundary artifacts mentioned in the qualitative assessment.

\begin{figure}[t]
\centering
\caption{Quantitative values for validation and test datasets of the BraTS Inpainting Challenge.}
\label{tab:challenge}
\vspace{0.5em}

\begin{tabular}{lcccccccc}
\toprule
\textbf{Dataset} &
\multicolumn{2}{c}{\textbf{SSIM $\uparrow$}} &
\multicolumn{2}{c}{\textbf{PSNR $\uparrow$}} &
\multicolumn{2}{c}{\textbf{MSE $\downarrow$}} &
\multicolumn{2}{c}{\textbf{RMSE $\downarrow$}} \\
\cmidrule(lr){2-3} \cmidrule(lr){4-5} \cmidrule(lr){6-7} \cmidrule(lr){8-9}
 & \textbf{Mean} & \textbf{Std} & \textbf{Mean} & \textbf{Std} & \textbf{Mean} & \textbf{Std} & \textbf{Mean} & \textbf{Std} \\
\midrule
Validation & 0.756 & 0.133 & 18.589 & 2.884 & 0.017 & 0.009 & 0.129 & 0.036\\
Test       & 0.786 & 0.145 & 17.737 & 3.469 & 0.019 & 0.015 & 0.139 & 0.056\\
\bottomrule
\end{tabular}

\end{figure}

\section{Discussion and Conclusion}
We present a unified neural network that performs 3D MRI inpainting for both brain tumors and healthy tissue. The model is conditioned on brain tissue segmentations and continuous tumor concentrations. Qualitative evaluation shows anatomically coherent reconstructions in every spatial direction and stable image quality at tissue boundaries, while the quantitative results confirm the model's strong performance.

\noindent Conditioning on tissue segmentations and tumor concentrations enables fine control over pathological and healthy tissue generation. The known region injection strategy preserves anatomical integrity in known areas while enabling robust inpainting in unknown ones.

\noindent Despite promising results, our system has certain limitations. Despite the advantages of latent diffusion, the training remains computationally intensive. Further, the repetitive repainting steps result in costly and time-consuming inference. Visual artifacts may still differentiate inpainted areas from original tissues in some outlier cases.

\noindent Future work will explore VAE fine-tuning, integration of additional MRI modalities, and dynamic tumor simulation for clinical decision support or tumor growth prediction. In the longer term, we aim to visualize an entire tumor trajectory by combining physics-based simulations with generative models, which enhances treatment planning, explainability and outcome forecasting.

\newpage
\bibliographystyle{splncs04}
\bibliography{references}
%
%
%
%




\end{document}